\documentstyle[prl,aps,epsf,two column]{revtex}

\include{psfig}

\title{Enhancement of  magnetic ordering  by the stress fields of
 grain boundaries in ferromagnets.
}  
\author{A. Kadigrobov*$^{1,2}$,
Z. Ivanov$^{1}$}
\address{
$^{1}$Department of Microelectronics and Nanoscience, Chalmers University of Technology and G\"oteborg University, 
S-412 96 G\"oteborg, Sweden\\
 $^{2}$B. I. Verkin Institute for Low Temperature Physics \& 
Engineering, \\
National Academy of Science of Ukraine, 47 Lenin Ave., 310164 Kharkov, 
Ukraine\\}
\author{R. I. Shekhter$^{3}$, and M.~Jonson$^{3}$
}
\address{
 $^{1}$Department of Applied Physics,
 Chalmers University of Technology and G\"oteborg University, 
SE-412 96 G\"oteborg, Sweden
}

\begin{document}

\date{}
\maketitle

\begin{abstract}
In the paper we predict a distinctive change of magnetic properties and considerable increase of the Curie temperature caused by the strain fields of grain boundaries in ferromagnetic films. It is shown that a sheet of spontaneous
magnetization  may arise along a grain boundary at temperatures greater than the bulk Curie temperature.  
The temperature dependence and space distribution of magnetization in a ferromagnetic film with grain boundaries  are calculated.  We found that $45^\circ$ grain boundaries  can produce long-range strain fields
that results in the width of the magnetic sheet along the boundary of the order of $ 0.5 \div 1 \mu m$ at temperatures grater than the bulk Curie temperature
by about  $10^2$ K.

\end{abstract}

Since discovery of the colossal magneto-resistance much attention has been payed to charge transport in magnetic materials keeping in mind their great potentiality for practical applications.
 In this regard  manganite perovskites are of a particular interest because  their magnetic and transport  properties  are strongly correlated as has been
observed  \cite{Hundley,Donnell} and explained on the base of  the double-exchange model proposed by Zener and de Gennese \cite{Zener,deGennes}. While in single crystals and high quality epitaxial films  of such materials magneto-resistance effects are large in strong magnetic fields of the order of 1$T$ close to the Curie temperature,  a large low field magneto-resistance has been established to arise  in thin films containing interfaces and grain boundaries \cite{Hwang,Steenbeck,Shreekala,Ziese,Lu,Isaac,Wang,Zdravko}. 
This low field effect appears due to electron spin polarized tunneling \cite{Lee,Gu,Guinea,Pin1,Pin2,Inoue}  or spin dependent scattering  at  grain boundaries/domain walls \cite{Wang}. 

Presence of grain boundaries in the sample  change not only its magneto-resistance but intrinsic magnetic properties itself   as it was observed in Ref.\cite{Demokritov,Zdravko}. In Ref.\cite{Zdravko}   grain boundaries were intentionally inserted 
in a film of $La_{0.7}Sr_{0.3}MnO_3$ that resulted in an increase of the ferromagnet transition temperature by more than  50 $K$ (an increase of the Curie temperature due to a strain caused by the grain boundaries was also reported    in \cite{Soh}).
 
In this paper we show that the strain field of a grain boundary 
 can result in a significant change of the magnetic structure of a ferromagnet giving rise to  a magnetic ordering phase in a large enough region  along the dislocation wall  at temperatures  noticeably higher than the bulk Curie temperature \cite{note1}.

Grain boundaries  in a crystal produce  strain fields and elastic deformations
that can increase the Curie temperature of a ferromagnet due to the dependence of the exchange 
energy on the distance between the neighboring atoms. This effect can be particularly large in $La_{1-x}Sr_{x}MnO_3$ materials where the ferromagnetic ordering is due to the double-exchange ferromagnetic coupling that is extremely sensitive to lattice distortion (see \cite{Blamire} and references there). 
 Hydrostatic pressure $p$ applied to $La_{1-x}Sr_{x}MnO_3$ ($0.15 \leq x \leq 0.5$)  increases  the Curie temperature  with a very high pressure coefficient 
\cite{Morimoto,Neumeier}:
\begin{equation}
\gamma = \partial  \ln T_c /\partial p \approx 0.065 GPa^{-1}
\label{experiment}   
\end{equation}

In this paper we assume  that  the grain boundary stress affects  ferromagnet properties
due to local change of the volume of the crystal as  ferromagnet parameters locally depend on the relative volume change (the elastic dilatation) $\epsilon_{ii}(x,y)$; ($\epsilon_{ik}$ - the strain tensor). 
In general, the influence of  local changes of ferromagnet parameters on the magnetization is not local 
due to long-range correlations in the ferromagnet
being   determined by the minimum condition of the ferromagnet free energy.
We 
find  the Curie temperature, the temperature dependence and space distribution of the equilibrium  magnetic moment ${\bf M} (x,y)$ of a ferromagnet with a tilt grain boundary solving the Landau-Lifshitz equation which is written as follows.  
\begin{equation}
-\alpha \frac{\partial^2 M}{\partial {\bf x}^2} + 2a\left(T - T_{c_0}-\delta T_{c}(x,y) \right)M 
+ 4 BM^3 =0
\label{moment}   
\end{equation}
Here $\alpha $ - the exchange constant, $a$ and $B$ are parameters in Ginsburg-Landau expansion of the free energy of a ferromagnet (see \cite{Landau}) ${\bf x}=(x,y,0)$ where $y$-axis is perpendicular to the grain boundary, $T_{c0}$ is the Curie temperature in the absence of the grain boundary; the local change of the critical temperature in Eq.(\ref{moment}) is 
\begin{equation}
 \delta T_c = - g T_{c0}\epsilon_{ii}(x,y), \;\; g=-\left .\frac{\partial \ln T_{c0} }{\partial \epsilon_{ii}}\right |_{\epsilon_{ii}=0} \approx K \gamma
\label{tempchange}
\end{equation}
where $K$ is the compression modulus.
The boundary conditions for Eq.(\ref{moment}) is  finiteness of $M(x,y)$  at the infinity \cite{note2}.

The dilatation   $\epsilon_{ii}(x,y)$, and hence the local temperature shift $\delta T_c(x,y)$ (see Eq.(\ref{moment},\ref{tempchange})) depends on the concrete structure of the boundary.

\begin{figure}
\centerline{\psfig{figure=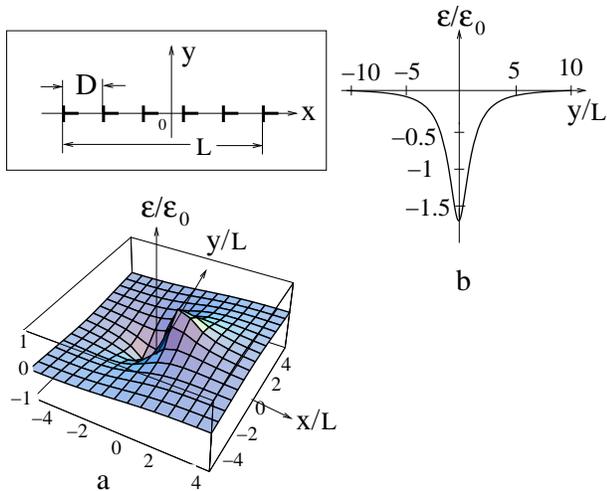,width=8cm}}
\vspace*{2mm}
\caption{a) Long-range relative dilatation produced by discontinuous tilt boundary of the length $L$ (shown in the insertion); b) dependence of the relative dilatation on the distance from the boundary at $x=1.2 L$.}
\label{Hlongrange}
\end{figure}

{\it  Boundaries of long-range stress fields.}
Despite the equilibrium grain  boundaries produce short-range stress fields 
around them, experiments and theory show that  boundaries of non-equilibrium configuration can do produce  long-range strain fields  in the crystal \cite{Hirth}. Below we present elastic dilatation for grain boundaries that may be relevant to the experimental situation of \cite{Zdravko,Soh}

a) For a  discontinuous tilt boundaries schematically shown in Fig.(\ref{Hlongrange}) the dilatation $\epsilon_{ii}$  is as follows \cite{Hirth}.
\begin{equation}
\epsilon_{ii}(x,y)= \frac{\epsilon_0}{2}\ln \frac{(x-L/2)^2 + y^2}{(x+L/2)^2 + y^2} 
\label{Hirth}   
\end{equation}
where
 $\epsilon_0 = (b/D)(1-2\sigma)/(1-\sigma)$, $D$ is the distance between
neighboring dislocations in the array, $\sigma$ is the Poisson ratio; $L\gg D$ is the length of the boundary. 
From Eq.(\ref{Hirth}) it follows that the elastic dilatation disappears at distances $\sim L \gg D$ (see also Fig.(\ref{Hlongrange})).

\begin{figure}
\centerline{\psfig{figure=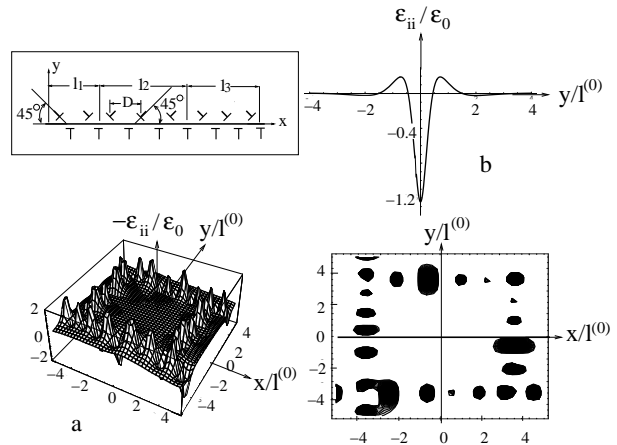,width=8cm}}
\vspace*{2mm}
\caption{a) Long-range relative dilatation produced by a non-equilibrium
 $45^\circ$ tilt boundary (shown in the insertion); b) dependence of the relative dilatation on the distance from the boundary at $x=-0.75 l_0$; c) magnetization regions along the boundary at temperature $(T-T_{c0})/T_{c0} = 0.1$}
\label{45longrange}
\end{figure}

b) {\it A $45^\circ$ tilt boundary.} In experiment \cite{Zdravko,Soh}  artificial boundaries in a cubic crystal were of such a structure that   on one side of the boundary ($y<0$) the $[100]$ crystal axis was parallel to the boundary while on the other side ($y>0$) the $[100]$ axis was rotated by $45^\circ$ with respect to the boundary. For the rotated part of the structure  ($y>0$) distortions in such a boundary  have two energetically equal directions due to the symmetry of the crystal with respect to the axis perpendicular the boundary plain. In the equilibrium configuration of the boundary the distortion in the boundary corresponds to only one of these two directions and long-range stress fields are absent. However, under the film growth  directions of the distortion
in different grains along the boundary  (nucleated at different points) may 
have different directions corresponding to the two  above-mentioned options
that results in a long-range stress field produced by such a boundary the width of which is of the order of the characteristic grain size. We find it using a misfit dislocation model of the boundary as is shown in Fig.(\ref{45longrange}).
Calculations show that in this case the dilatation $\epsilon_{ii}$
at distances much greater than the dislocation spacing  $D$ in the boundary ($\sqrt{x^2+y^2}\gg D$) is as follows. 
\begin{equation}
\epsilon_{ii}(x,y)=\epsilon_0 \sum^\infty_{n=-\infty}(-1)^n \ln\frac{(x-L_n -l_n)^2+y^2}{(x-L_n)^2+y^2}
\label{LR}   
\end{equation}
where $L_n = \sum_{k =0}^n l_n$
and $l_n$ is the length of the boundary  inside the $n$-th grain; these lengths are randomly distributed according to the random distribution of the 
sizes of the grains. In order to show its main feature we also present here dilatation 
$\epsilon_{ii}(x,y)$ for  a periodic structure of the boundary, assuming the period of it $l_0 \gg D$ to be the characteristic size of the grains  (that is $l_n =l_0 $); in this case the dilatation is as follows.
\begin{equation}
\epsilon_{ii}(x,y)=\epsilon_0 \ln\frac{1+2e^{(-\pi |x|/l_0)}\cos(\pi |y|) + e^{(-2 \pi |x|/l_0)}}{1-2e^{(-\pi |x|/l_0)}\cos(\pi |y| + e^{(-2 \pi |x|/l_0)}}
\label{lrperiod}   
\end{equation}
From Eq. (\ref{LR}) and Eq.(\ref{lrperiod}) one sees that such a boundary produces a long-range dilatation that spreads to distances $\sim l^{(0)}$ ($l^{(0)}$  - the characteristic size of the grains along the boundary, see Fig.(\ref{45longrange})).

According to Eq.(\ref{tempchange}) the dilatation Eq.(\ref{lowrange},\ref{Hirth},\ref{LR},\ref{lrperiod}) determine the local change of the Curie temperature $\delta T_c (x,y)$ in Eq.(\ref{moment}).
Inserting dimensional variables one sees the solutions of   Eq.(\ref{moment}) to be
governed by the following dimensionalless parameter  
\begin{equation}
\lambda = \left(\frac{d_0}{\xi_0}\right)^2
\frac{\delta T^{(0)}_c}{T_{c0}} \sim  \left(\frac{d_0}{\xi_0}\right)^2
K \gamma
\label{lambda}   
\end{equation}
where  $\delta T_c^{(0)}$ and  $d_0$ are the characteristic value and  the characteristic variation  in space of $\delta T_{c}(x,y)$, respectively;  $\xi_0=\sqrt{\alpha /(2a T_{c0})}$ is the characteristic correlation length in the ferromagnet at $T=0$.  
 We solve the  non-linear equation Eq.(\ref{moment}) for a general situation in two limiting cases $\lambda \ll 1$ and $\lambda \gg 1$ specifying  
 the  form of the  $\delta T_c (x,y)$  afterwards.

1) For the case $\lambda \ll 1$ we assume   $\epsilon_{ii}(x,y)$ to be localized along the boundary (decaying in the $y$-direction) and to be a periodic function in the $x$-direction with the period $d_0$. Under this assumption we  solve Eq.(\ref{moment}) in the following way. 

Expanding
the local critical temperature change  and the magnetization in the Fourier series 
\begin{equation}
\begin{array}{llc}
\delta T_c(x,y)/\delta T^{(0)}_c = \sum_{n=-\infty}^{\infty} V_n (y) \exp(i 2\pi n x/d_0),\\
m(x,y) = M(x,y)/M_0 =\\
\sum_{n=-\infty}^{\infty} A_n (y) \exp(i 2\pi n x/d_0)
\end{array}
\label{deltaT}   
\end{equation}
($M_0= \sqrt{a/2B}$ is the magnetization at $T=0$) and inserting it in  Eq.(\ref{moment}) one sees that
in the first non-vanishing approximation in $\lambda$, the  differential equation for any $A_n$ with $n \neq 0$ is a linear differential equations of the second order with constant coefficients and  the right side equal to $\lambda V_n(y)A_0(y)$ (that is they  are small comparing with $A_0$: $A_n(y)\sim \lambda A_0$). Solving these equations with the above-mentioned boundary conditions one gets 
the following non-linear equations for the zero harmonic $A_0$ of the magnetization $m$.  
\begin{equation}
 \frac{\partial^2  A_0 }{\partial \zeta^2}-\lambda \left(V_{eff}(\zeta) +  E \right)A_0 -C A_0^3 =0;
\label{A}   
\end{equation}
where 
$\zeta = y/d_0$, constant $C=(d_0/\xi_0)^2$,  "energy" $E=(T - T_{c0})/ \delta T_c^{(0)}$; the effective "potential" $V_{eff}(\xi)$ is as follows.
\begin{equation}
V_{eff}(\zeta)= -\lambda \sum_{n=1}^{\infty}\frac{1}{n}V_n(\zeta)\int_\infty^\infty V_n^{\ast}(\zeta+\zeta^{\prime})e^{-n \mid \zeta^{\prime} \mid} d\zeta^{\prime}
\label{eff}   
\end{equation}
While writing Eq.(\ref{A}) we used the relations $V_0 =0$, $V_n=-V_{-n}=-V_{n}^\ast$ which are valid for   the case of our interest $\delta T_c(x,y)=-\delta T_c(x,-y)$. 
   According to Eq.(\ref{A}), for $\lambda \ll 1$ function  $A_0(\zeta)$ varies at a distance that is  much grater than the interval where $V_{eff}(\zeta)$ is localized. Therefore, one may solve Eq.(\ref{A}) in the region $\zeta \gg 1$ (where $V_{eff}$
can be neglected) with the boundary condition 
\begin{equation}
\left . \frac{d A_0}{d \xi}\right|_{\xi=1} - \left . \frac{d A_0}{d \xi}\right|_{\xi=-1} = \lambda \int_{-\infty}^\infty V_{eff}(\xi) d\xi  A_0(0)
\label{boundA}   
\end{equation}

Non-linear Eq.(\ref{A}) with boundary condition Eq.(\ref{boundA}) has two bifurcation points: 1) at   at the bulk critical temperature $T=T_{c0}$, ($E=0$), and at a new critical temperature $T_c = T_{c0}+ \Delta T_c > T_{c0}$
($E= E_c = \Delta T_c/\delta T_c^{(0)}$)   at which a spontaneous magnetization arises around the grain boundary, where  
$\Delta T_c=  (\lambda \delta T_c^{(0)}/4)<V_{eff}>^2$.
Solving Eq.(\ref{A}) and Eq.(\ref{boundA}) one gets the magnetization to be as follows.
$$m(y)= \sqrt{2 (T_c -T)(T-T_{c0})/T_{c0}} )/$$
\begin{equation}
  \left(\sqrt{T-T_{c0}}\cosh (y/\xi_0(T)) +  \sqrt{Tc-T_{c0}} \sinh(|y|/\xi_0(T)\right)
\label{M1}   
\end{equation}
for $T_{c0} \leq T \leq T_{c}$, and
\begin{equation}
m(y) = \left(\frac{ T_{c0} - T }{T_{c0}}\right)^{1/2}\frac{1+a \exp\left(-\sqrt{2}|y|/\xi_0(T)\right)}{1-a \exp{\left(-\sqrt{2}|y|/\xi_0(T)\right)}  } 
\label {M2}  
\end{equation}
for $T < T_{c0}$, where 
\begin{equation}
 a= \sqrt{1 + 2 \frac{T_{c0}-T}{T_c -T_{c0}} }  -\sqrt{2 \frac{T_{c0}-T}{T_c -T_{c0}}}
\label{corrlength}   
\end{equation}
and the temperature dependent correlation length is $\xi_0 (T) = \xi_0 \sqrt{T_{c0}/|T-T_{c0}|}$

\begin{figure}
\centerline{\psfig{figure=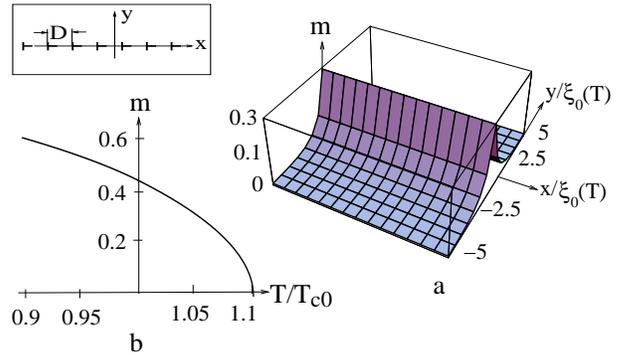,width=8cm}}
\vspace*{2mm}
\caption{a) Relative magnetization at $(T-T_{c0})/T_{c0} = 1.05$  and  $(T_c - T_{c0})/T_{c0} =1.1$ for $\lambda \ll 1$; b) temperature dependence of the magnetization at the boundary ($y=0$) }. The insertion shows  a low-angle tilt boundary. 
\label{Fig3}
\end{figure}

Therefore, for  $\lambda \ll 1$ the grain boundary enhanced magnetization in such a way that   in the temperature region $T_{c0} \leq T\leq T_C$ there is a sheet of magnetic ordering along the boundary the width of which is $\sim \xi_0(T)$ as shown in Fig.\ref{Fig3}

For example, the dilatation $\epsilon_{ii}(x,y)$ produced by  a low-angle tilt boundary (see the insertion in Fig.(\ref{Fig3}))  is \cite{Hirth})
\begin{equation}
\epsilon_{ii}(x,y) = -\epsilon_0 \frac{\sin (2 \pi x/D) }{\cosh (2 \pi y/D) - \cos(2 \pi x/D)}, 
\label{lowrange}   
\end{equation}
and  using Eq.(\ref{lowrange}, \ref{eff}) and Eq.(\ref{deltaT}) one finds   $ <V_{eff}> = -1.72 \lambda$ and the new Curie temperature  $T_c$ to be
\begin{equation}
T_c = T_{c0}(1 - 0.43 \lambda^3 g \epsilon_0) \approx T_{c0}(1+ 0.43 \gamma \lambda^3 K \epsilon_0)
\label{lowrangetemp}   
\end{equation}

For the case $\lambda \gg 1$  the pattern of 
the magnetization  temperature dependence  differs   and may be analyzed  as follows.

The Curie temperature $T_c$ is found as the lowest level $E_0$ of  the  "energy" $E = T-T_{c0}$ of the linearized  Eq.(\ref{moment}) \cite{Nabutovski}.
 For $\lambda \gg 1$ it can be easily found because the eigenfunction corresponding
to this lowest level should have no nodal lines \cite{Morse} while
 wave functions of the linearized Eq.(\ref{moment}) oscillate at distances
$\delta r \sim \lambda^{-1} d_0$. It means that  the lowest level should be  
so close to the bottom of the "potential well" that the ground state function
is localized around the point of the minimum of it ${\bf x}_0= (x_0, y_0)$ at distances $|{\bf x} - {\bf x}_0| \sim \lambda^{-1} d_0 \ll d_0$ (where only one oscillation of the ground state wave function takes place), and hence the Taylor series expansion of  $\delta T_c (x,y)$ in Eq.(\ref{moment}) is possible. In this case calculations show that  magnetic ordering appears in a narrow tube of the width $\sim \lambda^{-(1/4) }d_0$ around   ${\bf x}_0$ at the temperature $T=T_c + \delta T_c$ where  the shift of the critical temperature is    
$\delta T_c \approx T_{c0}|g| \epsilon_0 \sim K \gamma \epsilon_0 T_{c0}$.

With a further decrease of temperature the region of the magnetization expands 
and for $T_c < T < T_{c0}$,  according to Eq.(\ref{moment}), the magnetization of the ferromagnet in the presence of the boundary can be written as  
\begin{equation}
m (x,y) = \sqrt{1 -  g \epsilon_{ii}(x,y) -T/T_{c0}}
\label{lrm}   
\end{equation}
 As follows from Eq.(\ref{lrm}) and Eq.(\ref{Hirth}, \ref{LR}, \ref{lrperiod}), 
in the regions where $\epsilon_{ii}< 0$,
spontaneous magnetization arises along the boundary at distances of the order
of the grain size at temperatures exceeding the bulk temperature $T_{c0} $  (see Fig.(\ref{45longrange})). With further
decrease of the temperature, in the temperature interval where $\xi_0(T)\gg l_0$, the magnetization is described by Eq.(\ref{M1}, \ref{M2}) (see also Fig.(\ref{Fig3}) with $d_0 = l_0$.

 Using  Eq. (\ref{tempchange}) and Eq. (\ref{experiment},\ref{LR})
one estimates the increase of the Curie temperature due to the long-range boundary strain to be $\delta T_c^{(0)}/T_{c 0}\approx \gamma K (b/D)$ ($K$ is the compressibility). For the experimental value of $\gamma = 0.065 GPa$ \cite{Morimoto,Neumeier}, $T_{c0}$ =350 K, $D \approx 10 b$, and  typical values $K = 50 GPa$
one gets $\delta T_c^{(0)} \approx 100$K while the characteristic 
size of the regions of the spontaneous magnetization at $T_{c0} < T < T_{c0}+\delta T_c^{(0)}$ being $l^{(0)}\sim 0.1 \div 1 \mu m$ for the grain size typical for  the experiment. \vspace{2 mm}

A.K acknowledges the hospitality of the Department of Applied Physics and the Department of Microelectronics and Nanoscience,  Chalmers University of Technology and G\"oteborg University.

The study was supported by SSF OXIDE  and SSF QUANTUM DEVICES and NANOSCIENCE programs.\vspace{3 mm}

*E-mail: kadig@fy.chalmers.se~and~kadig@kam.kharkov.ua


\begin{thebibliography}{99}

\bibitem{Hundley} F.M. Hundley et al., Appl. Phys. Lett. {\bf 67}, 806 (1995).

\bibitem{Donnell} J. O'Donnell et al., Phys. Rev. B {\bf 54}, 6841 (1996).


\bibitem{Zener} C. Zener,  Phys. Rev.  {\bf 81}, 440 (1951).

\bibitem{deGennes} P.G. de Gennes,  Phys. Rev. {\bf 118}, 141 (1960).

\bibitem{Hwang} H.Y. Hwang et al., Phys. Rev. Lett., {\bf 77}, 2041 (1996).

\bibitem{Steenbeck} K. Steenbeck et al., Appl. Phys. Lett. {\bf 71}, 968 (1997);
K. Steenbeck et al., ibid. {\bf 73}, 2506 (1998).

\bibitem{Shreekala} R. Shreekala et al., Appl. Phys. Lett. {\bf 71}, 282 (1997).

\bibitem{Ziese} M. Ziese et al., Appl. Phys. Lett. {\bf 74}, 1481 (1997); M. Ziese,  Phys. Rev. B {\bf 60}, R738 (1999).

\bibitem{Lu} Yu Lu et al.,  Phys. Rev. B {\bf 54}, R8357 (1996); X.W. Li et al., J. Appl. Phys. {\bf 81}, 5509 (1997)

\bibitem{Isaac} S.P. Isaac et al.,  Appl. Phys. Lett. {\bf 72}, 2038 (1998).

\bibitem{Wang} H.S. Wang and Qi Li, Appl. Phys. Lett. {\bf 73}, 2360 (1998); H.S. Wang et al., ibid. {\bf 74}, 2212 (1999) 

\bibitem{Demokritov} S. Demokritov, U. R\"{u}cker, P. Gr\"{u}nberg, J. Magn. Magn. Mater. {\bf 163}, 21 (1996). 

\bibitem{Zdravko}  R. Mathieu, P. Svedlindth, R.A. Chakalov, Z.G. Ivanov, Phys. Rev.B {\bf 62}, 3333 (2000)


\bibitem{Lee} S. Lee, H. Hwang, B.I. Shraiman, W.D. Ratcliff II, and S-W Cheong, Phys. Rev. Lett. {\bf 82}, 4508 (1999).

\bibitem{Gu} R.Y. Gu, D.Y. Xing, and Jinming Dong, J. Appl. Phys. {\bf 80},
7163 (1996).

\bibitem{Guinea} F. Guinea, Phys. Rev. B {\bf 58}, 9212 (1998).

\bibitem{Pin1} Pin Lyu, D.Y. Xing, and Jinming Dong,  Phys. Rev. B {\bf 58}, 55 (1998).

 \bibitem{Pin2} Pin Lyu, D.Y. Xing, and Jinming Dong,  Phys. Rev. B {\bf 60}, 4235 (1999).

\bibitem{Inoue} J. Inoue and S. Maekava, J. Magn. Magn. mater. {\bf 198 - 199}, 167 (1999).

\bibitem{Soh} Eeong-Ah Soh, G. Aeppli, N.D. Mathur, and M.G. Blamire,  Phys. Rev. B {\bf 63}, 020402(R)) 
 
\bibitem{note1}  We note here that an increase
of the   local critical temperature for a  superconductor in the strain field of a dislocation was predicted in Ref. \cite{Nabutovski}; an increase of the superconductor local critical temperature  for  HTSC materials due to long-range strain fields of the dislocation arrays of crystalline defects was studied in Ref.\cite{Pashitski}.

\bibitem{Nabutovski} N.M. Nabutovskii and V. Ya. Shapiro, Sov. Phys. JETP, {\bf 48}, 480 (1978).

\bibitem{Pashitski} A. Gurevich and E.A. Pashitskii,  Phys. Rev. B {\bf 56}, 6213 (1997).

\bibitem{Blamire} M.G. Blamire et al., J. Magn. Magn. Mater. {\bf 191}, 359 (1998)

\bibitem{Morimoto} Y. Moritomo and A. Asamitsu, Phys. Rev. B {\bf 51}, 16491 (1995).

\bibitem{Neumeier} J.J. Neumeier, M.F. Hundley, J.D. Thomson, and R.H.Heffner,
 Phys. Rev. B {\bf 52}, R7006 (1995).

\bibitem{note2} Such an equation for a single dislocation and a dislocation wall in a superconductor was solved in Ref.\cite{Nabutovski} and  Ref.\cite{Pashitski}, respectively.  We  solve Eq.(\ref{moment}) using another method which  permits to analyze an effect of a strain field on the ferromagnet magnetization   in wider ranges of parameters in more details without referring to specific properties of dislocations.


\bibitem{Landau} L.D. landau, E.M. Lifshits {\it Electrodynamics of Continuous Media},
Chapter V, Pergamon Press, 1993.


\bibitem{Hirth} J.P. Hirth, J. Lothe, {\it Theory of Dislocations}, Part 4
(McGraw-Hill, New York, 1968)

\bibitem{Morse} P.M. Morse and H. Feshbach, Methods of Theoretical Physics,
Vol.1, Chap. VI, McCrow-Hill Book Company, 1953.



\end{thebibliography}
\end{document}